\documentclass{article}

\usepackage{PRIMEarxiv}

\usepackage[utf8]{inputenc} 
\usepackage[T1]{fontenc}    
\usepackage{hyperref}       
\usepackage{url}            
\usepackage{booktabs}       
\usepackage{amsfonts}       
\usepackage{amsmath,bm}
\usepackage{nicefrac}       
\usepackage{microtype}      
\usepackage{lipsum}
\usepackage{fancyhdr}       
\usepackage{graphicx}       
\usepackage{rotating,tabularx}
\usepackage{multirow}
\usepackage{algpseudocode}

\usepackage{bbm}
\usepackage{subcaption}
\usepackage{dsfont}

\usepackage{amsthm}
\usepackage[ruled,vlined]{algorithm2e}

\theoremstyle{definition}

\theoremstyle{remark}

\newtheorem{theorem}{Theorem}[section]

\newcommand{\vect}[1]{\boldsymbol{#1}}

\DeclareMathOperator*{\argmax}{arg\,max}

\graphicspath{{media/}}     

\title{Parametric estimation of conditional Archimedean copula generators for censored data
\thanks{\textit{\underline{Citation}}: 
\textbf{Michaelides, M., Cossette, H. \& Pigeon, M., Parametric estimation of conditional Archimedean copula generators for censored data (2024). Preprint.} 
}}

\author{
  Marie Michaelides\\
  Chaire Co-operators en analyse des risques actuariels \\
  Département de mathématiques \\
  Université du Québec à Montréal \\
  Montréal\\
  \texttt{michaelides.marie@uqam.ca} \\
  \And
  Mathieu Pigeon \\
  Chaire Co-operators en analyse des risques actuariels \\
  Département de mathématiques \\
  Université du Québec à Montréal \\
  Montréal\\
   \And
  Hélène Cossette \\
  École d'actuariat \\
  Université Laval \\
  Québec\\
}

\begin{document}
\maketitle

\begin{abstract}
In this paper, we propose a novel approach for estimating Archimedean copula generators in a conditional setting, incorporating endogenous variables. Our method allows for the evaluation of the impact of the different levels of covariates on both the strength and shape of dependence by directly estimating the generator function rather than the copula itself. As such, we contribute to relaxing the simplifying assumption inherent in traditional copula modeling. We demonstrate the effectiveness of our methodology through applications in two diverse settings: a diabetic retinopathy study and a claims reserving analysis. In both cases, we show how considering the influence of covariates enables a more accurate capture of the underlying dependence structure in the data, thus enhancing the applicability of copula models, particularly in actuarial contexts.
\end{abstract}

\section{Introduction}\label{sec:introduction}
Copulas are a powerful tool when we want to address the dependence among correlated variables. Widely used since their introduction by \cite{sklar1959} in various fields from finance and insurance to medicine, they allow to separate the dependence structure within a multivariate distribution. Multiple authors have extensively studied their different families. The Archimedean copulas, introduced by \cite{schweizer1983}, are particularly useful thanks to their easy characterization via a generator function $\psi(.)$. To model the dependence for a vector of $d$ variables $\vect{Y} = (Y_1,...,Y_d) \in \mathbb{R}^d$, the $d-$variate Archimedean copula $C(.)$ takes the form
\begin{align*}
    C(F_1(y_1),...,F_d(y_d)) = \psi\big\{\psi^{-1}[F_1(y_1)] + ... + \psi^{-1}[F_d(y_d)]\big\},
\end{align*}
where the generator function $\psi(.):[0,\infty] \to [0,1]$ must be such that $\psi(0)=1$ and $\lim_{\nu \to \infty}\psi(\nu) = 0$. From a copula $C(.)$, one can recover Kendall's tau and as such, assess the degree of association between the variables under consideration. Archimedean copulas are typically characterized by a small number of dependence parameters, some of them even possessing only one for which a direct relationship exists with Kendall's tau. This is notably the case for the Clayton, Frank, Gumbel and Joe copulas. 

Although widely popular in the literature for a few decades now, it is only in recent years, thanks to the early work of \cite{patton2006}, that some authors have began to consider the impact of risk factors on the dependence between random variables. \cite{patton2006} extended Sklar's theorem to incorporate a covariate $Z$ in the copula model. 
More specifically, this novel approach allows to model the dependence between the random variables $(Y_1,...,Y_d)$, conditionally on a covariate $Z$. The conditional copula $C(.|Z=z)$ is given by the joint distribution function of $(F_{1|Z}(Y_1|z),...,F_{d|Z}(Y_d|z))$, where $Z=z$. In the bivariate case, \cite{patton2006} showed that the joint conditional distribution is then uniquely defined for any $z$ in the support of $Z$ as
\begin{align*}
    H_Z(y_1,y_2|z) = C\big(F_{1|Z}(y_1|z), F_{2|Z}(y_2|z) | Z=z\big),
\end{align*}
for all $(y_1,y_2) \in \mathbb{R}^2$.

Very often in conditional copula models, authors rely on the so called \textit{simplifying assumption}. This assumption states that a copula $C(.|Z)$ is independent of $Z$: 
\begin{align*}
    C\big(F_{1|Z}(y_1|z), F_{2|Z}(y_2|z) | Z = z\big) = C\big(F_{1|Z}(y_1|z), F_{2|Z}(y_2|z) \big) \hspace{0.3cm}\forall z \in Z.
\end{align*}
The effects of the covariate $Z$ are only captured through the marginal distributions $F_{1|Z}(.)$ and $F_{2|Z}(.)$ while the copula itself remains unchanged for all values of $z$. The simplifying assumption is particularly useful in pair copula constructions that allow to specify multivariate distributions using only bivariate copulas as building blocks. They allow to capture highly complex dependence structures. In addition, assuming that the conditional copulas used in the hierarchical structure do not depend on the conditioning variables keeps the resulting multivariate distribution tractable for inference and model selection. Although this assumption greatly simplifies conditional copula models, voices have been raised in the statistical literature against it, see for example \cite{hobaek2010}, \cite{acar2012} or \cite{stober2013}. In particular, some authors like \cite{levi2018} argue for the need of effective tests against the simplifying assumption.  

Some contributions to the literature have worked towards relaxing this assumption. Most of these focus on incorporating endogenous variables in the strength of dependence, often to estimate Kendall's tau or, equivalently, the dependence parameters for Archimedean copulas, as in \cite{acar2011}. However and to the best of our knowledge, no model presented so far allows for both the strength and the structure of dependence to implicitly vary with the values of a covariate.


The approach that we present in this paper aims at relaxing the simplifying assumption in the case of Archimedean copulas. More specifically, we propose a new parametric estimator for the generator function $\psi(.)$, suitable for censored data and inspired by the non-parametric estimator presented in \cite{genest1993}. With this new parametric model, we include covariates in the estimation of the generator function, thereby considering their impact on both the strength and shape of dependence between random variables of interest. This contrasts with most approaches presented in the literature that incorporate endogenous variables directly in the copula $C(.)$ rather than in the generator function. Besides enabling the relaxation of the simplifying assumption, this approach is parametric, contrarily to a lot of the models in the existing literature. Although non-parametric frameworks help avoiding assumptions that could be misleading, incorporating covariates in these models often proves difficult, and interpreting their effects is seldom straightforward. By using parametric models, we can more easily include multiple covariates and more readily analyze their impact. Another interesting feature of our model is that it is suitable for various censoring schemes for the dependent random variables. Incomplete data is often encountered in a variety of fields and although multiple authors have proposed copula models that can accommodate different types of censoring, see for example \cite{lopez2015}, very few so far allow to incorporate endogenous variables. Finally, while most contributions to conditional copula models are found in the medical field, this paper contributes to the actuarial literature. To the best of our knowledge, very few authors have used such frameworks in actuarial modeling. Considering the important quantity of information available to insurers, incorporating risk factors related to the claims or the policyholders in actuarial models helps refine the predictions made by actuaries. In this paper, we present an application of a conditional copula model to micro-level loss reserving. 

This paper is structured as follows. In Section \ref{sec:review}, we review the existing conditional copula models as well as some tests for the simplifying assumption. Section \ref{sec:model} presents the parametric estimator for the generator function of Archimedean copulas. In Section \ref{sec:RDS}, we apply our model to the Diabetic Retinopathy dataset used, among others, by \cite{geerdens2018}. We present an application to granular reserving in Section \ref{sec:insurance}. Using a Canadian automobile insurance dataset in which each policy in force provides different coverages, we investigate the dependence between these coverages and the impact that some covariates may have on this dependence. Section \ref{sec:conclusion} concludes our work.

\section{Literature review}
\label{sec:review}
In this section, we present an overview of the literature on conditional copulas. We discuss the contributions made since the pioneering work of \cite{patton2006}, starting with the parametric approaches before moving on to the non-parametric and semi-parametric models. We then discuss some of the works related to the simplifying assumption. 

As mentioned in Section \ref{sec:introduction}, \cite{patton2006} was the first to set the basis of conditional copula modeling by extending Sklar's theorem. He uses conditional copulas to capture the dependence between the Deutsche mark-dollar and Yen-dollar exchange rates and the effect of time on this dependence, particularly before and after the introduction of the Euro. Several applications of these conditional copulas emerged in the financial literature simultaneously, each concentrating on integrating time-varying aspects into the dependence structure of ARMA models.

Several authors followed suit, marking the start of the literature on conditional copulas. Most contributions to date fall in the non-parametric domain. Among the first, \cite{acar2011} use likelihood estimation for covariate-adjusted copulas in a local polynomial framework. Assuming the conditional marginal distributions $F_{1|Z}$ and $F_{2|Z}$ known, the authors focus on the model
\begin{align*}
    (U_{1i},U_{2i}) | Z_i \sim C(u_{1i}, u_{2i}|\theta(z_i)),
\end{align*}
with $\theta(z_i) = g^{-1}\big\{ \eta(z_i)\big\}$, for $i=1,..,n$, the dependence parameter. Here, $g^{-1}$ is a known inverse link function that keeps the dependence parameter in its appropriate range. $\eta$ is the unknown calibration function that the authors estimate using local maximum likelihood and the framework of local polynomials. 

In a similar vein, \cite{valle2017} also opt for a non-parametric approach, using a flexible Bayesian framework for the estimation of conditional copula densities. Their method allows to overcome the issue of selecting among various types of copulas. \cite{bouezmarni2019} use an inverse-probability-of-censoring weighting approach in a bivariate case to derive a non-parametric estimator of the conditional marginal distribution with one covariate when only one of the correlated variables is subject to random censoring. Focusing solely on modeling the strength of dependence, \cite{derumigny2019} directly estimate the conditional Kendall's tau in the presence of a vector of covariates by using different kernel-based methods. The authors then prove several theoretical properties of these estimators. In \cite{derumigny2022}, the same authors prove the weak convergence of conditional empirical copulas processes under various conditioning events with non-zero probabilities, extending the work of \cite{segers2012}. 

In the actuarial literature, only few authors until now have delved into the early work of \cite{patton2006} and explored conditional copulas within actuarial models. Among rare examples, \cite{yang2020} propose a non-parametric method to estimate copula regression models with discrete outcomes, introducing a perturbed version of the probability integral transform. They apply their methodology to estimate insurance claim frequencies across different business lines. Also in the context of multiple business lines in insurance, but this time in a claims reserving setting, \cite{yang2022} develops a copula estimator for mixed data. In his paper, the author accounts for the mixed nature of claims severities with a probability mass at zero and a positive continuous part. 

Although the literature on conditional copula accounts for a lot of non-parametric models, some authors have also built on the work of \cite{patton2006} to propose parametric approaches, mostly using Bayesian regression copulas. This is the case, for example, in \cite{pitt2006}, with different types of outcomes and distributional assumptions for the marginals. \cite{fermanian2012} propose time-dependent copulas where each dependent random variable depends on its value at the previous lag. Another interesting example, \cite{hans2022} use boosting to estimate the coefficients of Generalized Additive Models for the Location, Scale and Shape (GAMLSS) for the parameters of the marginal distributions and dependence parameters of Gaussian, Clayton and Gumbel copulas. The authors apply their methodology to birth cohort data to predict the correlated birth length and birth weight of newborns, using $36$ covariates. Recently, \cite{wei2023} propose to model time to graft failure and time to death since kidney transplantation with conditional survival copulas using parametric distributions for the marginals. They then evaluate the effects of different covariates by means of hazard ratios estimated from the copula models. One of the few examples handling censored data, \cite{geerdens2018} use a local likelihood approach for the copula parameter with both parametrically and non-parametrically estimated marginal survival functions. The authors also propose a generalized likelihood ratio test to evaluate the constancy of the conditional copula parameter.

Most applications of these Bayesian regression copulas are, however, found in medicine, and only a few examples have found their way to the actuarial literature. Notably, \cite{shi2022} employ a copula regression framework with vine copulas to jointly estimate a policyholder' deductible, claim frequency and claim amounts. 

Other propose models in-between the non-parametric and parametric frameworks. \cite{klein2021} use a copula decomposition of the joint distribution of a vector of values from a single response variable. The authors assume that the dependence structure is an unknown smooth (parametric) function of the covariates and use non-parametric estimators for the marginal distributions. They construct the copula by inverting a pseudo regression. \cite{liu2021} propose a semi-parametric conditional mixture copula model, where the copula is a weighted average of different conditional copulas. They model not only the copula parameters but also the weights of each individual copula in the mixture based on a covariate and using a two-step semi-parametric estimation procedure. This allows to model highly flexible dependence structures. Very recently in the actuarial literature, \cite{wang2023} apply conditional copulas to model the dependence between a pair of discrete and continuous outcomes, namely the number of claims and average claim amount observed for policyholders. They use the distribution regression approach to provide a semi-parametric estimation framework for the joint distribution of these outcomes. More precisely, they use regressions to estimate the marginal distribution of the number of claims using covariates. Knowing this, they then estimate the marginal distribution for the claim amount, conditional on the covariates and on the number of claims. Combining these two regressions, they are able to retrieve the joint conditional distribution.

 Most contributions mentioned in this section so far rely on the simplifying assumption, i.e. endogenous variables are included in the model solely through the marginal distributions and they assume that the copula does not vary with the covariates. As mentioned in Section \ref{sec:introduction}, some authors have however challenged this assumption and proposed formal tests to assess its validity in recent years. A nice overview can be found in \cite{derumigny2017}. Among others, \cite{gijbels2017} develop a test based on a Rao-type score statistic. Their test allows both for assessing the impact of a vector of risk factors on the dependence parameter but also on the specification of the copula model. Working with three-dimensional vine copulas, \cite{czado2017} investigate the differences between simplified and non-simplified models. \cite{spanhel2019} question the simplifying assumption for pair copula constructions and explore cases where it does not hold, leading to a partial vine copula approximation. In \cite{kurz2022}, the same authors present a test for the simplifying assumption, applicable to high-dimensional vine copulas by discretizing the support of the conditioning covariate and using a penalty in the proposed test statistic.

\section{Parametric estimator for Archimedean copulas generators}
\label{sec:model}
In this section, we present a parametric model to estimate the generator function of Archimedean copulas in the presence of endogenous variables. We show how directly working with the generator rather than the copula itself relaxes the simplifying assumption by allowing both the dependence parameters and shape of dependence to vary with different levels of the covariates.

We consider the vector of times-to-events $\vect{T}=(T_{1},T_{2})$. Note that we present here the bivariate case for the sake of simplicity. Let $\vect{X}=(X_{1},X_{2})$ be the vector of censoring times such that we only observe 
\begin{align*}
    \vect{Y} &= (Y_{1},Y_{2}) =  \big( \min\{T_{1},X_{1}\}, \min\{T_{2},X_{2}\} \big)
\end{align*}
and the censoring indicators 
\begin{align*}
    \delta_{j} = \mathds{1}_{[T_{j} \leq X_{j}]},
\end{align*}
for $j=1,2$. Let $\vect{Z} \in \mathbb{R}^{n\times P}$ be a vector of risk factors, where $\vect{Z}_{i} = (Z_{i,1},Z_{i,2},...,Z_{i,P})$ is the vector of risk factors for the $i^{\text{th}}$ observation. 

Our goal is to capture the dependence between $Y_1$ and $Y_2$ with an Archimedean copula, while including the effect of the covariates $\vect{Z}$. To do this, we consider the following non-parametric estimator of the Archimedean generator function $\psi^{-1}(.)$ first introduced by \cite{genest1993}: 
\begin{align}
\label{eq:generator}
    \psi^{-1}(\nu) = \exp \Bigg\{\int_{\nu_0}^\nu \frac{1}{t-K(t)}dt \Bigg\},
\end{align}
with $0<\nu_0<1$ an arbitrarily chosen constant and where $K(\nu)$ is the univariate Kendall distribution, defined as 
\begin{align*}
    K(\nu) = \nu - \frac{\psi^{-1}(\nu)}{\psi^{{-1}^{(1)}}(\nu)}, \hspace{0.3cm} 0<\nu \leq 1.
\end{align*}
In the presence of censored data, \cite{wang2000} propose the following non-parametric estimator for the Kendall distribution
\begin{align}
\label{eq:kendall}
    \hat{K}_n(\nu) = \int_0^\infty \int_0^\infty \mathds{1}_{[\hat{F}(\vect{y}) \leq \nu]}d\hat{F}(\vect{y}),
\end{align}
where $\hat{F}(\vect{y})$ is an appropriate non-parametric estimator for the joint distribution of $Y$. A non-parametric estimator for the Archimedean generator function $\hat{\psi}^{-1}_n(.)$ can then be recovered by substituting (\ref{eq:kendall}) in (\ref{eq:generator}).

When both $T_1$ and $T_2$ are subject to random censoring, \cite{akritas2003} propose the following estimator for the joint distribution
\begin{align}
\label{eq:joint}
    \hat{F}(\vect{y}) = w(\vect{y})\int_0^{y_2} \hat{F}_{1 \vert 2}(y_1  \vert  \xi_2)d\Tilde{F}_2(\xi_2) + (1 - w(\vect{y}))\int_0^{y_1} \hat{F}_{2 \vert 1}(y_2  \vert  \xi_1)d\Tilde{F}_1(\xi_1),
\end{align}
where $\Tilde{F}_{1}$ and $\Tilde{F}_{2}$ are the marginal estimators of Kaplan and Meier (1958): 
\begin{align*}
	\Tilde{F}_{j}(y_{i}) = 1 - \prod_{Y_{i,j} \leq y_{i}, \Delta_{ij}=1} \Big(1 - \frac{1}{n-i+1}\Big), \hspace{0.3cm}j=1,2,
\end{align*}
and the weights $w(\vect{y})$ minimize the mean-squared error of $\hat{F}(\vect{y})$. The functions $\hat{F}_{1 \vert 2}(y_1  \vert  y_2)$ and, similarly, $\hat{F}_{2 \vert 1}(y_2  \vert  y_1)$, are the estimators of the conditional distributions of, respectively, $Y_1$ and $Y_2$. These are obtained using the following extension of \cite{beran1981}'s estimator:
\begin{align}
\label{eq:beran}
    \hat{F}_{1 \vert 2}(y_1  \vert  y_2) = 1 - \prod_{Y_{i,1} \leq y_{1}, \Delta_{i,1} = 1} \Bigg( 1 - \frac{W_{ni2}(y_{2};h_{n})}{\sum_{j=1}^{n}W_{nj2}(y_{2};h_{n})\mathds{1}_{Y_{j,1} \geq Y_{i,1}}}\Bigg).
\end{align}
In this estimator, the values of $y_2$ must be uncensored and
\begin{align*}
	W_{ni2}(y_2;h_{n}) = \begin{cases}  \frac{k\Big(\frac{y_2-Y_{i,2}}{h_{n}}\Big)}{\sum_{\Delta_{j,1}=1}k\Big(\frac{y_2-Y_{j,2}}{h_{n}}\Big)},&\hspace{0.3cm}\text{if}\hspace{0.2cm} \Delta_{i2}=1 \\ 0,&\hspace{0.3cm}\text{if}\hspace{0.2cm} \Delta_{i2}=0,  \end{cases}
\end{align*}
where $k(.)$ is a known kernel function and $\{h_n\}$ is the bandwidth: a sequence of positive constants such that $h_n \rightarrow 0$ as $n \rightarrow \infty$. A similar estimator is used for the conditional distribution of $Y_2$.

Although using Beran's estimator allows to incorporate a covariate in the model, the method suffers from a lack of interpretability of the effect of this covariate on the response. In addition, including more than one covariate can become tedious. We propose instead a parametric model for the conditional distribution of $Y_1$ (resp. $Y_2$) given $Y_2$ (resp. $Y_1$) and the vector of continuous or discrete covariates $\vect{Z}$. More specifically, we use the framework of generalized additive models for location, scale and shape (GAMLSS) to build censored regressions for both $Y_1$ and $Y_2$. The great flexibility of GAMLSS allows us to choose the best fitting distribution among a very large selection and to estimate the different parameters of the chosen distribution using linear, non-linear or smooth functions of the covariates. As such, we are able to build and compare various censored parametric models and better assess the impact of each covariate on the conditional distributions. 

We thereby seek to model $(Y_{i,j} \vert \vect{Z}_i = \vect{z}_i, Y_{i,k}=t_{i,k})$, the observed time-to-event for observation $i$, with $i=1,...,n$ and $j=1,2$, knowing the vector of risk factors $\vect{Z}_i$ and the second time-to-event $t_{i,k}, k\neq j$. Given a candidate distribution $F_j(.|\vect{\theta}_j)$ with a vector of $D$ parameters $\vect{\theta}_j$, we use GAMLSS as follows:
\begin{align}
\label{eq:gamlss}
    \theta_{i,j,d} = g^{-1}(\vect{z}_i' \vect{\beta}_{p,j} + \beta_{j}^* t_{i,k}),
\end{align}
where $g(.)$ is an appropriate link function, $\vect{\beta}_{p,j}$ is a $1\times P$ vector of coefficients associated to covariate $Z_p$ and linked to parameter $\theta_{j,d}$ that is estimated, and $\beta_{j}^*$ is the specific coefficient attached to the (uncensored) activation delay of coverage $k$, for $k=1,2$ and $k\neq j$ and observation $i$. Following \cite{geerdens2018}, one can easily retrieve the coefficient estimates via maximum likelihood using
\begin{align}
\label{eq:beta}
    \hat{\vect{\beta}}_{j,d} = \argmax \sum_{i=1}^{n} \delta_{i,j} \ln f_{j|\vect{Z},t_k}(y_{i,j}| \vect{Z}_i ,t_{i,k}, \vect{\beta}_{p,j}) + (1-\delta_{i,j})\ln F_{j|\vect{Z},t_k}(y_{i,j}| \vect{Z}_i, t_{i,k}, \vect{\beta}_{p,j}),
\end{align}
where $f(.)$ and $F(.)$ are, respectively, the density and cumulative distribution functions of the chosen parametric model. 

Once we select an appropriate model for $Y_j$, we estimate the model coefficients for the different distribution parameters using (\ref{eq:beta}) and then obtain the parameters estimates $\hat{\vect{\theta}}_j$ for each observation $i$ using (\ref{eq:gamlss}). We can now easily obtain the estimators for the conditional distributions
\begin{align}
\label{eq:cond}
    \hat{F}_{j|\vect{Z},t_k}(y_j | \vect{Z}, t_k, \hat{\vect{\theta}}_j).
\end{align}
For a $1 \times P$ vector of covariates, the estimation of the joint distribution then becomes 
\begin{align}
\begin{split}
\label{eq:jointnew}
    \hat{F}(\vect{y}) = w(&\vect{y}) \int_0^{z_1} \hdots \int_0^{z_P} \int_0^{y_2} \hat{F}_{1|\vect{Z},t_2}(y_1 | \vect{\zeta},\xi_2,\hat{\vect{\theta}}_1)d\Tilde{F}_{2|\vect{Z}}(\xi_2|\vect{\zeta}) dF_{Z_P}(\zeta_P)\hdots dF_{Z_1}(\zeta_1) \\
    &+ (1 - w(\vect{y})) \int_0^{z_1}\hdots \int_0^{z_P} \int_0^{y_1} \hat{F}_{2|\vect{Z},t_1}(y_2 | \vect{\zeta}, \xi_1,\hat{\vect{\theta}}_2)d\Tilde{F}_{1|\vect{Z}}(\xi_1|\vect{\zeta})dF_{Z_P}(\zeta_P) \hdots dF_{Z_1}(\zeta_1).
\end{split}
\end{align}
From (\ref{eq:jointnew}), it is then possible to isolate the effects of one covariate, or of a subset of covariates, to assess their impact on the joint distribution. Suppose that we want to analyze the impact of the subset $\vect{Z}^Q = (Z_1, ..., Z_Q) \subset \vect{Z}$, with $Q < P$. We then have  
\begin{align}
\begin{split}
\label{eq:jointnew2}
    \hat{F}(\vect{y}| &\vect{Z}^Q) = w(\vect{y}) \int_0^{z_{Q+1}} \hdots \int_0^{z_P} \int_0^{y_2} \hat{F}_{1|\vect{Z},t_2}(y_1 | \vect{\zeta},\xi_2,\hat{\vect{\theta}}_1)d\Tilde{F}_{2|\vect{Z}}(\xi_2|\vect{\zeta}) d\Breve{F}_{Z_P}(\zeta_P)\hdots d\Breve{F}_{Z_{Q+1}}(\zeta_{Q+1}) \\
    &+ (1 - w(\vect{y})) \int_0^{z_{Q+1}}\hdots \int_0^{z_P} \int_0^{y_1} \hat{F}_{2|\vect{Z},t_1}(y_2 | \vect{\zeta}, \xi_1,\hat{\vect{\theta}}_2)d\Tilde{F}_{1|\vect{Z}}(\xi_1|\vect{\zeta})d\Breve{F}_{Z_P}(\zeta_P) \hdots d\Breve{F}_{Z_{Q+1}}(\zeta_{Q+1}).
\end{split}
\end{align}
In (\ref{eq:jointnew}) and (\ref{eq:jointnew2}), $\Tilde{F}_{j|\vect{Z}}(y_j|\vect{Z})$ for $j=1,2$ are the marginal parametric estimators, fitted through a GAMLSS, and $\Breve{F}_{Z_p}(Z_p)$ for $p=1,...,P$ are the marginal estimators for each covariate. 

With this new parametric estimator for the joint distribution $\hat{F}(\vect{y})$, we propose the following estimator for the univariate Kendall distribution: 
\begin{align*}
    \hat{K}(\nu) = \int_0^{\infty} \int_0^{\infty} \mathds{1}_{[\hat{F}(\vect{y}) \leq \nu]}d\hat{F}(\vect{y}) = \nu - \hat{\lambda}(\nu),
\end{align*}
where $\hat{F}(\vect{y})$ is the estimator shown in (\ref{eq:jointnew}). Similarly as before, we can isolate the effects of a subset of covariates $\vect{Z}^Q \subset \vect{Z}^P$, with $1 \leq Q \leq P$, using
\begin{align}
\label{eq:Kcov}
    \hat{K}(\nu | \vect{Z}^Q) = \int_0^{\infty} \int_0^{\infty} \mathds{1}_{[\hat{F}(\vect{y}) \leq \nu]}d\hat{F}(\vect{y}| \vect{Z}^Q) =  \nu - \hat{\lambda}(\nu | \vect{Z}^Q).
\end{align}
Thanks to the estimator shown in (\ref{eq:Kcov}), we obtain a different Kendall distribution based on the values of the endogenous variables of interest. Finally, we obtain a parametric estimator for the Archimedean copula generator function:
\begin{align}
\label{eq:psi}
    \hat{\psi}^{-1}(\nu) = \exp \Bigg\{\int_{\nu_0}^\nu \frac{1}{t-\hat{K}(t)}dt \Bigg\},
\end{align}
or, with the isolated effects of a subset of covariates,
\begin{align}
\label{eq:psicov}
    \hat{\psi}^{-1}(\nu | \vect{Z}^Q) = \exp \Bigg\{\int_{\nu_0}^\nu \frac{1}{t-\hat{K}(t | \vect{Z}^Q)}dt \Bigg\}.
\end{align}
Thanks to the estimator of the Kendall distribution in (\ref{eq:Kcov}), we can compute Kendall'a tau either across all covariates, as shown below
\begin{align*}
    \hat{\tau} = 4 \int_0^1 \hat{\lambda}(\nu)d\nu = 3 - 4\int_0^1 \hat{K}(\nu)d\nu,
 \end{align*}
or while taking into account the impact of the selected subset of risk factors, using the relation
 \begin{align}
\label{eq:taucov}
    (\hat{\tau} | \vect{Z}^Q) = 4 \int_0^1 \hat{\lambda}(\nu | \vect{Z}^Q)d\nu = 3 - 4\int_0^1 \hat{K}(\nu| \vect{Z}^Q)d\nu.
 \end{align}
Even more interestingly, in addition to being able to quantify the impact of a subset of covariates on the strength of dependence as in (\ref{eq:taucov}), using the estimator proposed in (\ref{eq:psicov}) allows the risk factors to impact the generator function and thereby, the shape of dependence. As such, we can relax the simplifying assumption when using $\hat{\psi}^{-1}(\nu | \vect{Z}^Q)$ since we do not work with the hypothesis that the structure of dependence is invariant to the effects of the covariates.

\section{Diabetic Retinopathy Study}
\label{sec:RDS}
To illustrate the methodology laid out in Section \ref{sec:model}, we consider the Diabetic Retinopathy Study (RDS) dataset used by \cite{geerdens2018} and \cite{huster1989}. This dataset contains 197 entries, corresponding to 197 diabetic patients, for which we observe the effect of a treatment by laser photocoagulation in delaying the time to blindness. For each patient, one eye is randomly selected to remain untreated while we observe the effect of the laser treatment in the other eye. The objective is to model the dependence between the time to blindness of both eyes, as well as the impact of a covariate, namely the age at onset of diabetes, on this dependence. The times to blindness, in months, can be censored if blindness did not occur while the patient was observed in the study. 

For each entry $i$ in the dataset, for $i=1,...,197$, we observe the following vector: $(Y_{i,1}, Y_{i,2}, \delta_{i,1}, \delta_{i,2}, Z_{i})$. $Y_{i,1}$ (resp. $Y_{i,2}$) is the time to blindness for the treated (resp. untreated) eye, $\delta_{i,1}$ (resp. $\delta_{i,2}$) is the censoring indicator for the treated (resp. untreated) eye, equal to $0$ if the observation is censored, i.e. if the study has stopped before the onset of blindness in that eye. The time to blindness in the treated eye is censored for $73\%$ of the 197 patients observed in the dataset, against only $49\%$ for the untreated eye.  
The covariate $Z_{i}$ represents the age at onset of diabetes (in years) for the individual, ranging from 1 to 58 years old.

\subsection{Dependence modeling without covariate}
We first analyze the dependence structure present in this dataset without taking the covariate $Z$ into account. Assuming that we can model the data with an Archimedean copula, we find an estimator of the generator function $\psi^{-1}(.)$ and use it to estimate Kendall's tau. We compare two approaches: first, we find the estimator $\hat{\psi}^{-1}_n(\nu)$ non-parametrically, combining the method proposed in \cite{genest1993} in (\ref{eq:generator}) and \cite{beran1981}'s estimators for the conditional marginal distributions, as in \cite{akritas2003} in (\ref{eq:joint}). We use Epanechnikov kernels and select the optimal bandwidth by cross-validation for Beran's estimators. Second, we replace these non-parametric estimators by parametric models for the marginal distributions of the times to blindness. As such, we estimate the joint distribution using (\ref{eq:jointnew}), without the terms related to the vector of covariates $\vect{Z}$. Following \cite{geerdens2018}, we opt for Weibull marginal distributions, with parameter vectors $\vect{\theta}_j = (\theta_{j,1}, \theta_{j,2})$ for $j=1,2$, where $\theta_{j,1}$ and $\theta_{j,2}$ are, respectively, the scale and shape parameters for the time to blindness in eye $j$. The joint distribution is then given by 
\begin{align*}
    \hat{F}(\vect{y}) = w(\vect{y})\int_0^{y_2} \hat{F}_{1 \vert 2}(y_1  \vert  \xi_2)d\Tilde{F}_2(\xi_2) + (1 - w(\vect{y}))\int_0^{y_1} \hat{F}_{2 \vert 1}(y_2  \vert  \xi_1)d\Tilde{F}_1(\xi_1),
\end{align*}
where 
\begin{align*}
    \hat{F}_{1 \vert 2}(y_{i,1}  \vert  \xi_{i,2}) = 1 - \exp\Big[-\Big(\frac{y_{i,1}}{\hat{\theta}_{i,1,1}}\Big)^{\hat{\theta}_{i,1,2}} \Big],
\end{align*}
with 
\begin{align*}
    \hat{\theta}_{i,1,1} = \exp(\hat{\beta}_{0,1} + \hat{\beta}^*_{1} \xi_{i,2}),
\end{align*}
the scale parameter for the Weibull distribution, where $\hat{\beta}_{0,1}$ is the intercept of the model for parameter $\hat{\vect{\theta}}_1$ and $\hat{\beta}^*_1$ is the coefficient linked to $\xi_2$. For the shape parameter, we have 
\begin{align*}
    \hat{\theta}_{i,1,2} = \exp(\hat{\gamma}_{0,1} + \hat{\gamma}^*_{1} \xi_{i,2}),
\end{align*}
where $\hat{\gamma}_{0,1}$ is the intercept of the model for parameter $\hat{\vect{\theta}}_2$ and $\hat{\gamma}^*_1$ is the coefficient linked to $\xi_2$. $\Tilde{F}_2(y_2)$ is the parametric marginal estimator of $y_2$. We proceed similarly for $\hat{F}_{2 \vert 1}(.)$.

\begin{figure}[h!]
\centering
\includegraphics[width=1\textwidth]{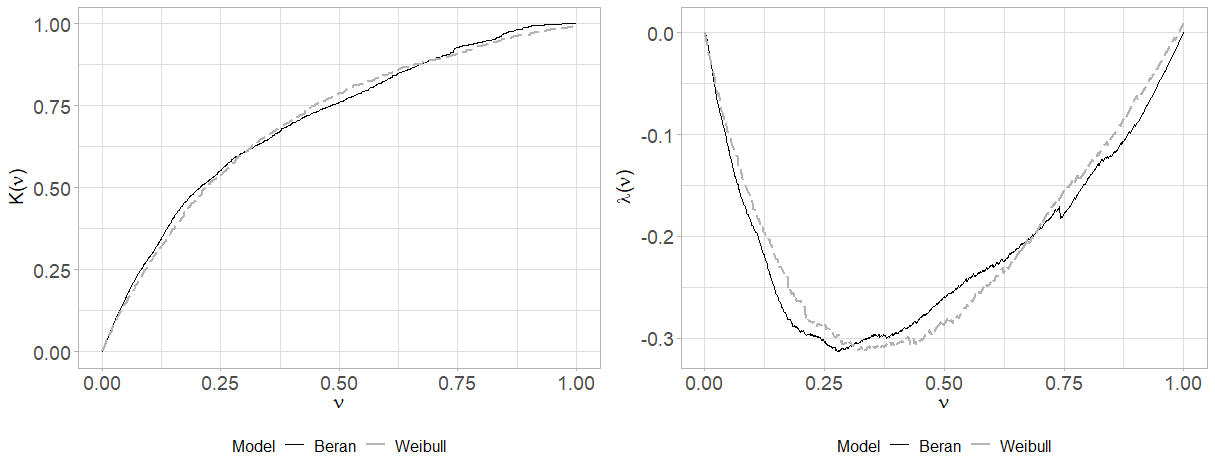}
\caption{$\hat{K}(\nu)$ and $\hat{\lambda}(\nu)$ with the non-parametric (black lines) and parametric (gray dashed lines) approaches for the RDS data.}
\label{fig:RDS}
\end{figure}

Figure \ref{fig:RDS} shows the estimated Kendall distribution $\hat{K}(.)$ (left) and, equivalently, $\hat{\lambda}(.)$ (right), obtained using both the non-parametric (black curves) and parametric (grey curves) approaches from, respectively, (\ref{eq:generator}) and (\ref{eq:psi}). 

We observe that the results of both methods are very close. The curves of the $\hat{\lambda}(.)$ functions on the right plot display similar shapes, indicating that the dependence structure is the same with both estimations, which was to be expected. The resulting values for Kendall's tau are $\hat{\tau}_{\text{NP}} = 0.1864$ and $\hat{\tau}_{\text{P}} = 0.1859$ for, respectively, the non-parametric and the parametric models. The estimated strength of dependence is thus also almost identical. Since we do not include any risk factor in the model at this stage, we expect both approaches to give similar results. 

\subsection{Dependence modeling with a covariate}
We now enter the age at onset of diabetes as covariate in the parametric model. The joint distribution thus becomes
\begin{align*}
\begin{split}
    \hat{F}(\vect{y}) = w(\vect{y}) \int_0^{z} \int_0^{y_2} \hat{F}_{1|Z,t_2}(y_1 | \zeta, &\xi_2)d\Tilde{F}_{2|Z}(\xi_2|\zeta) dF_Z(\zeta)  \\
    &+ (1 - w(\vect{y})) \int_0^{z} \int_0^{y_1} \hat{F}_{2|Z,t_1}(y_2 | \zeta, \xi_1,)d\Tilde{F}_{1|Z}(\xi_1|\zeta)dF_Z(\zeta),
\end{split}
\end{align*}
with 
\begin{align*}
    \hat{F}_{1 \vert Z,t_2}(y_{i,1}  \vert  \zeta_i, \xi_{i,2}) = 1 - \exp\Big[-\Big(\frac{y_{i,1}}{\hat{\theta}_{i,1,1}}\Big)^{\hat{\theta}_{i,1,2}} \Big],
\end{align*}
where
\begin{align*}
    \hat{\theta}_{i,1,1} = \exp(\hat{\beta}_{0,1} + \hat{\beta}_{1,1} \xi_{i,2} + \hat{\beta}^*_1 \zeta_{i})
\end{align*}
and
\begin{align*}
    \hat{\theta}_{i,1,2} = \exp(\hat{\gamma}_{0,1} + \hat{\gamma}_{1,1} \xi_{i,2} + \hat{\gamma}^*_1 \zeta_{i}).
\end{align*}
For $\hat{F}_{2|Z,t_1}$, we have
\begin{align*}
    \hat{F}_{2 \vert Z,t_1}(y_{i,2}  \vert  \zeta_i, \xi_{i,1}) = 1 - \exp\Big[-\Big(\frac{y_{i,2}}{\hat{\theta}_{i,2,1}}\Big)^{\hat{\theta}_{i,2,2}} \Big],
\end{align*}
where
\begin{align*}
    \hat{\theta}_{i,2,1} = \exp(\hat{\beta}_{0,2} + \hat{\beta}_{1,2} \xi_{i,1} + \hat{\beta}^*_2 \zeta_{i})
\end{align*}
and
\begin{align*}
    \hat{\theta}_{i,2,2} = \exp(\hat{\gamma}_{0,2} + \hat{\gamma}_{1,2} \xi_{i,1} + \hat{\gamma}^*_2 \zeta_{i}).
\end{align*}

\begin{figure}[h!]
\centering
\includegraphics[width=1\textwidth]{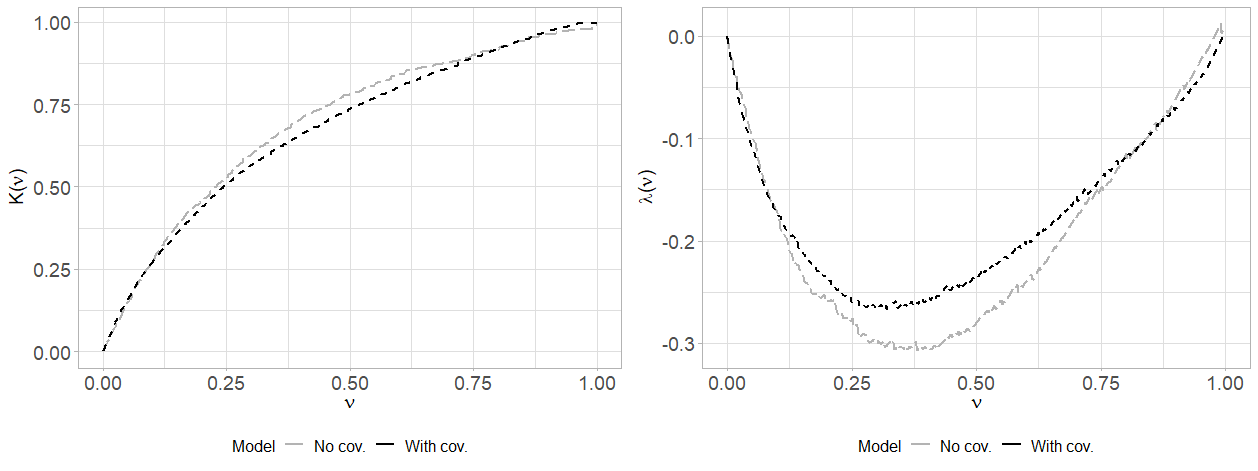}
\caption{$\hat{K}(\nu)$ and $\hat{\lambda}(\nu)$ with (black dotted lines) and without (gray dashed lines) covariate for the RDS data, under the parametric model.}
\label{fig:RDScov}
\end{figure}

Figure \ref{fig:RDScov} shows again the curves of the estimated Kendall distribution on the left and of $\hat{\lambda}(\nu)$ on the right, this time for the parametric model with (black dotted line) and without (grey dashed lines) the covariate. We observe a clear jump in the curve, corresponding to a jump in the value of Kendall's tau, from $\hat{\tau} = 0.1859$ in the model without the age at onset of diabetes to $\hat{\tau} = 0.3001$ when including this variable in the joint distribution of the times to blindness. 

This risk factor thus has an important impact on the strength of dependence when considering all the individuals in the dataset. We now investigate the impact of different values of the covariate on the dependence model. Using (\ref{eq:jointnew2}), we 
could obtain a separate estimator of the generator function for each different value of the age at onset of diabetes, that is a different curve for the Kendall distribution for each age. For illustrative purposes, we analyze the impact for individuals who were younger and older than 20 years old at onset of diabetes. The choice of this splitting point is, at this stage, only motivated by the results derived in \cite{geerdens2018}, where the authors show that with the inclusion of the covariate in the model, the average estimate of Kendall's tau for the whole dataset is around $0.3$. It is below (resp. above) this value for individuals who were first diagnosed with diabetes before (resp. after) the age of 20. This splitting point also separates the dataset into two subsets of roughly the same size. We will further discuss the choice of the splitting point for the conditioning variable in Section \ref{sec:conclusion}.

\begin{figure}[h!]
\centering
\includegraphics[width=1\textwidth]{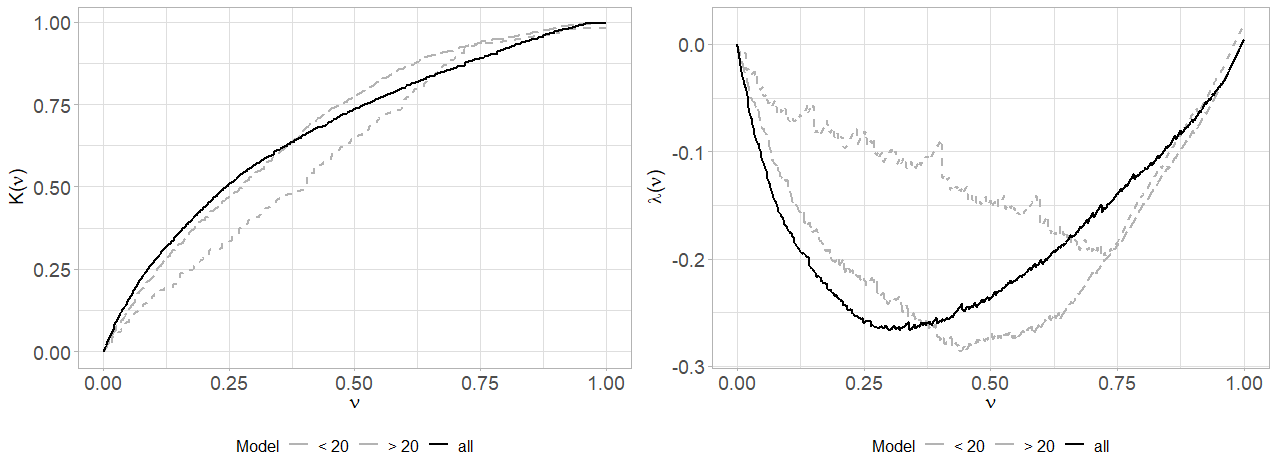}
\caption{Plots of $\hat{K}(\nu)$ (black continuous curve), $\hat{K}(\nu | Z \leq 20)$ (grey dashed curve) and $\hat{K}(\nu | Z > 20)$ (grey dotted curve) on the right and $\hat{\lambda}(\nu)$, $\hat{\lambda}(\nu | Z \leq 20)$ and $\hat{\lambda}(\nu | Z > 20)$ on the left.}
\label{fig:RDScovlevels}
\end{figure}

Figure \ref{fig:RDScovlevels} compares the curves of the Kendall distribution and the lambda function for all individuals, i.e. $\hat{K}(\nu)$, (in black) and for those who were younger or older than twenty at onset of diabetes (in grey), i.e. respectively, $\hat{K}(\nu | Z \leq 20)$ and $\hat{K}(\nu | Z > 20)$. The black curve for all individuals is the same as the one depicted in Figure \ref{fig:RDScov}. Interestingly, we observe that the shapes of the three curves on both plots are quite different, indicating that conditionally on the value of the covariate, both the strength and the shape of dependence vary. 

We first show the differences on the strength of dependence via Kendall's tau in Table \ref{tab:taucomp}. While the estimated value for the population as a whole is around $30\%$, it it slightly smaller at $26.3\%$ for individuals who were first diagnosed with diabetes before turning twenty, and quite higher at almost $56\%$ for individuals who started suffering from diabetes after the age of twenty. These results are in line with those presented in \cite{geerdens2018}.

\begin{table}[h]
\begin{center}
\caption{Kendall's tau for different values of the age covariate $Z$}
\label{tab:taucomp}
\begin{tabular}{@{}c c@{}}
\toprule
Model & $\hat{\tau}$ \\
\midrule
All &  0.3001371 \\
$ Z \leq 20$ & 0.2629644 \\
$ Z> 20$ & 0.5591597 \\
\bottomrule
\end{tabular}
\end{center}
\end{table}

Next, we observe the impact of these two different age groups on the shape of dependence. From Figure \ref{fig:RDScovlevels}, the shape of the $\hat{\lambda}(.)$ curve for the dataset as a whole points towards the bottom left corner of the plot, depicting the typical shape of a Gumbel or a Joe copula. The curve for the patients whose disease started before the age of 20 points more towards the center or the bottom right corner of the plot, indicating that a Frank or a Clayton copula might be more fitted for this age group. To determine which model best fits these two populations, we perform simulations from the estimated generator functions and apply the $L^2-$norm model validation method used in, among others, \cite{lakhal2010}. We use the \texttt{RPLATRANS} algorithm described in \cite{ridout2009}, together with the Marshall, Olkin algorithm from \cite{marshall1988}. The \texttt{RPLATRANS} algorithm allows to sample from a distribution specified by its inverse Laplace-Stieltjes transform. Knowing this, we can use the Marshall, Olkin algorithm to simulate bivariate vectors of observations $(U_1,U_2)$ from the Archimedean copula from which the data originates. More details on these algorithms are provided in Appendix \ref{app:1}. More specifically, we perform the simulation procedure presented below.

\begin{algorithm}
\caption{Copula simulation and selection procedure}\label{alg:sim}
\begin{algorithmic}[1]
        \item For $j$ in $1,...,J$ where $J$ is the total number of simulations performed, simulate a new bivariate sample of size $n$ from the parametric estimator $\hat{\psi}(\nu)$, using the \texttt{RLAPTRANS} and Marshall, Olkin algorithms: $(\vect{Y}^{(j)}_1, \vect{Y}^{(j)}_2) = \{(Y_{1,1}^{(j)},Y_{1,2}^{(j)}),...,(Y_{n,1}^{(j)},Y_{n,2}^{(j)})\}$.
        \item For each new simulated sample $(j)$, compute again the Kendall distribution $\hat{K}^{(j)}(\nu)$ and estimate $\hat{\tau}^{(j)}$.
        \item For the $M$ candidate copula models under consideration, use $\hat{\tau}^{(j)}$ to estimate the dependence parameter $\hat{\alpha}_m$, for $m=1,...,M$. For each candidate model, estimate $K_{\hat{\alpha}_m}^{(j)}(\nu)$. 
        \item Calculate the $L^2-$norm between $K_{\hat{\alpha}_m}^{(j)}(\nu)$ and $\hat{K}^{(j)}(\nu)$ as 
        \begin{align*}
            D^{(j)}(\hat{\alpha}) = \int_\xi^1 \big( \hat{K}^{(j)}(\nu) - K_{\hat{\alpha}_m}^{(j)}(\nu)\big)^2 d\nu.
        \end{align*} 
        \item For each candidate model $m$, obtain the pseudo $p-$value as
        \begin{align*}
            p_m = \frac{1}{J}\sum_{j=1}^{J}\mathds{1}[\min_l D^{(j)}(\hat{\alpha}_{l}) > D^{(j)}(\hat{\alpha}_{m})],
        \end{align*}
        with $l, m = 1,...,M$ and $l \neq m$.
\end{algorithmic}
\end{algorithm}
In Algorithm \ref{alg:sim}, the pseudo $p$-value $p_m$ can be interpreted as the percentage of simulated samples for which candidate model $m$ presents the smallest distance between its Kendall distribution and that of the data. The copula with the largest pseudo $p-$value will be the most appropriate model. 

\begin{figure}[h!]
\centering
\includegraphics[width=1\textwidth]{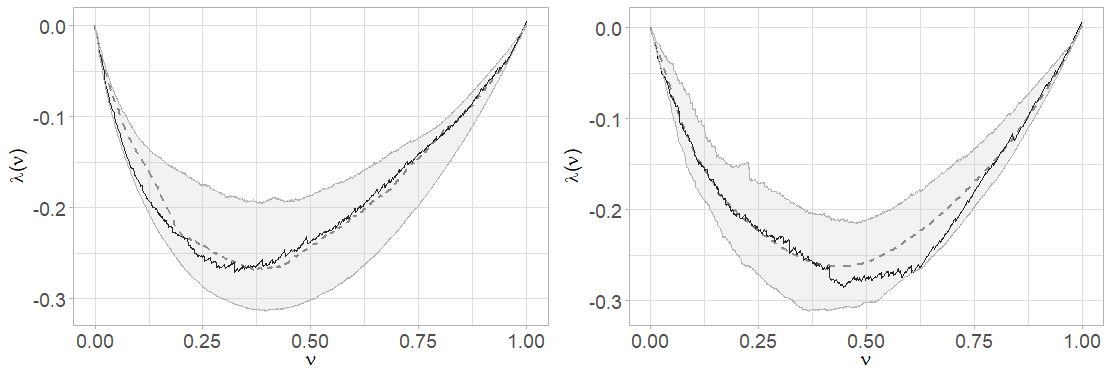}
\caption{$\lambda(.)$ for the samples simulated using the generator functions for the whole population (left) and for the patients who were first diagnosed with diabetes before the age of twenty (right).}
\label{fig:allvsyoung}
\end{figure}

Figure \ref{fig:allvsyoung} shows the results of $J=1000$ simulations for the whole dataset (left) and for the individuals who had diabetes before the age of twenty. The black curves correspond to the $\hat{\lambda}(\nu)$ estimated parametrically from the original data. The dashed grey curves represent the average of the simulations and the shaded areas show the $95\%$ confidence intervals. 

Table \ref{tab:L2} shows which copula models among the Clayton, Frank, Gumbel and Joe were selected with the validation procedure for each simulation. Note that the results are presented here in terms of (1-$p$-values). The best fitted model is then the one with the smallest value displayed in the table. 
For the subset of individuals whose diabetes started before the age of twenty, 460 out of 1000 simulations select the Clayton copula as the most appropriate model, 305 simulations prefer the Frank copula and 235 opt for the Gumbel model. None of the simulations result in the Joe copula being selected as best fitted to this subset of the data. 
For the data as a whole, only 138 out of the 1000 simulations select the Clayton copula as best fitted. Most simulations, that is 464 out of the 1000, opt for the Frank model, 354 select the Gumbel copula and 41 choose the Joe model. The Frank copula thus appears to be best fitted for the whole data, followed by the Gumbel copula while for the subset of patients who were diagnosed with diabetes before turning twenty, the Clayton copula is the preferred model, followed by the Frank copula in second place. The age at onset of diabetes thereby seems to have an impact not only on the strength of dependence as discussed earlier, but also on the shape of dependence. When modeling the correlated times to blindness in both eyes, one might want to use different copula models for different groups of patients. 

Note, however, that the pseudo $p$-values in Table \ref{tab:L2} do not lead to a clear decision, particularly for the Clayton, Frank and Gumbel that present relatively similar values. More specifically, with the high levels of right-censoring observed in the data, copulas such as the Clayton and Frank may more closely resemble each other, leading to uncertainty in the choice of the most appropriate model. In their paper, although \cite{geerdens2018} only use the covariate for modeling the dependence parameter, they implement a generalized likelihood ratio test to determine whether the assumed copula model is adequate. The test leads them to similar conclusions as observed here. However, since our approach leads to an estimator of the copula generator, we do not need to specify a known copula family, contrarily to other approaches found in the literature so far. Our method allows us to conclude that the conditioning variable impacts both the level and shape of dependence between the times to blindness. Even if we can not define specific copula families with a sufficient level of confidence for different levels of the covariate, we can directly work from the estimated generator functions obtained in the different cases and as such, acknowledge that the copula model varies with the conditioning variable.

\begin{table}[h]
\begin{center}
\caption{Results from the copula selection procedure using the $L^2$-norm.}
\label{tab:L2}
\begin{tabular}{@{}c c c c c@{}}
\toprule
$Z$ & Clayton & Frank & Gumbel & Joe \\
\midrule
$\leq 20$ & 0.540 & 0.695 & 0.765 & 1.000 \\
All & 0.862 & 0.536 & 0.646 & 0.959 \\
\bottomrule
\end{tabular}
\end{center}
\end{table}

\section{Automobile insurance dataset application}
\label{sec:insurance}
In this section, we apply the non-parametric estimator for Archimedean copula generators presented in Section \ref{sec:model} to a loss reserving context, using a Canadian automobile insurance dataset. 

The data includes over $600~000$ claims that occurred between 2015 and 2021. Each of these claims relate to a policy in force under which four insurance coverages are provided: the Accident Benefits, Bodily Injury, Vehicle Damage and Loss of Use coverages. Note that, to simplify the presentation, we will work in a bivariate framework and only consider the Accident Benefits and Bodily Injury coverages that are the most important cost-wise in the portfolio. 

From its moment of declaration to the insurer, each new claim will impact at least one of these coverages. We define the \textit{activation delay} of a claim for a coverage as the time elapsed between the moment of declaration of the claim and the first moment at which the insurer realizes that this claim triggers the coverage and registers this information in his claims management system. For illustrative purposes, consider a claim declared on June $1^{\text{st}}$, 2016. On the same day, based on the information available at the time, the insurer labels it as an Accident Benefits claim. The activation delay for this claim under this coverage is then equal to one day. On June $25^{\text{th}}$, the insurer receives new information leading to label the claim as a Bodily Injury claim as well. The activation delay for this claim under the Bodily Injury coverage is thereby 25 days. 

Considering that not all claims impact all coverages, the activation delays are subject to right-censoring. This occurs when, for a given claim that has reached its settlement date, a coverage has not been triggered yet. The censoring variable is thus the settlement delay, defined here as the time elapsed between the declaration and settlement dates of a claim. In this portfolio, $21.63\%$ of all claims impact both the Accident Benefits and Bodily Injury coverages. These are the claims for which neither of the activation delays are censored. For $23.46\%$ of claims, only the Bodily Injury coverage is triggered. This corresponds to the percentage of claims for which we observe a censored value for the Accident Benefits coverage. The remaining $54.91\%$ of claims only activate the Accident Benefits delay, i.e. these are the claims for which the activation delays of the Bodily Injury coverage are censored.   

\begin{figure}[h!]
\centering
\includegraphics[width=.4\textwidth]{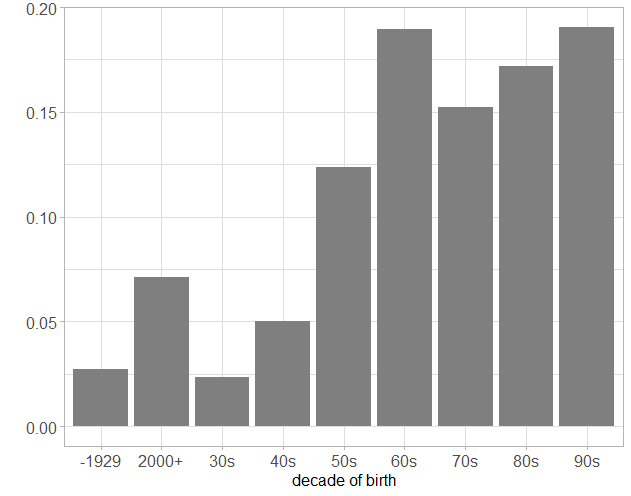}
\caption{Frequency of the different decades of birth of individuals in the dataset.}
\label{fig:YOB}
\end{figure}

In addition to the coverages, the dataset provides for each claim information related to the policyholder, the vehicle driven or the claim itself. Our goal is to model the dependence between the coverages provided within a single policy by means of their activation delays and the impact of some risk factors on this dependence, using the framework of Archimedean copulas. More specifically, we focus on the decade of birth of the policyholders, which is provided as a covariate in the dataset. Figure \ref{fig:YOB} shows more information about this risk factor. We choose to create a new categorical variable with three levels, grouping the different decades of birth in three groups of approximately equal sizes. Let $Z_i$ be the value of this new covariate for claim $i$, such that 
\begin{align*}
    Z_i = \begin{cases}
        1, &\hspace{0.3cm}\text{if the policyholder was born in or after 1990} \\
        2, &\hspace{0.3cm}\text{if the policyholder was born between 1970 (incl.) and 1989} \\
        3, &\hspace{0.3cm}\text{if the policyholder was born in or before 1969}.
    \end{cases}
\end{align*}

We will use this covariate as the conditioning variable in our copula model to investigate the possible effects of the age of policyholders on the strength and structure of dependence between the activation delays. We use the following notation: 
\begin{itemize}
    \item $\vect{T}_i = (Y_{i,1}, Y_{i,2})$ is the vector of activation delays for our two coverages, for claim $i$ with $i=1,...,n$.
    \item $\vect{X}_i = (X_{i,1}, X_{i,2})$ is the vector of censoring times for both coverages and for claim $i$. This vector corresponds in this case to the settlement delay of the claim.
    \item $\vect{Y}_i = (Y_{i,1}, Y_{i,2}) = (\min(T_{i,1}, X_{i,1}), \min(T_{i,2}, X_{i,2}))$ is the vector of observed activation delays for claim $i$, taking censoring into account.
    \item $\delta_{i,j} = \mathds{1}_{(T_{i,j}\leq X_{i,j})}$ for $j=1,2$ are the censoring indicators for claim $i$.
    \item $Z_i$ is the age of the policyholder upon occurrence of claim $i$, belonging to one of three groups previously defined.
\end{itemize}
For illustration purposes, suppose that a policyholder born in the 1980s files a claim following a car accident. His claim only triggers the Accident Benefits coverage upon reporting, i.e. with a delay of one day, and settles 90 days later, without activating further coverages. The data entry related to this claim $i$ is $(Y_{i,1},Y_{i,2},\delta_{i,1},\delta_{i,2},Z_i) = (1, 90, 1, 0, 2)$.

\subsection{Dependence modeling without covariate}
Like we did for the diabetic retinopathy study, we first model the dependence between the censored activation delays without conditioning on the covariate. We compare the results derived from a non-parametric model using Beran's estimators in the joint distribution to those derived from a parametric model where we replace Beran's estimators by censored parametric regressions. Similarly to what we did in Section \ref{sec:RDS}, we choose Weibull distributions to model the marginal distributions of both activation delays. 

\begin{figure}[h!]
\centering
\includegraphics[width=1\textwidth]{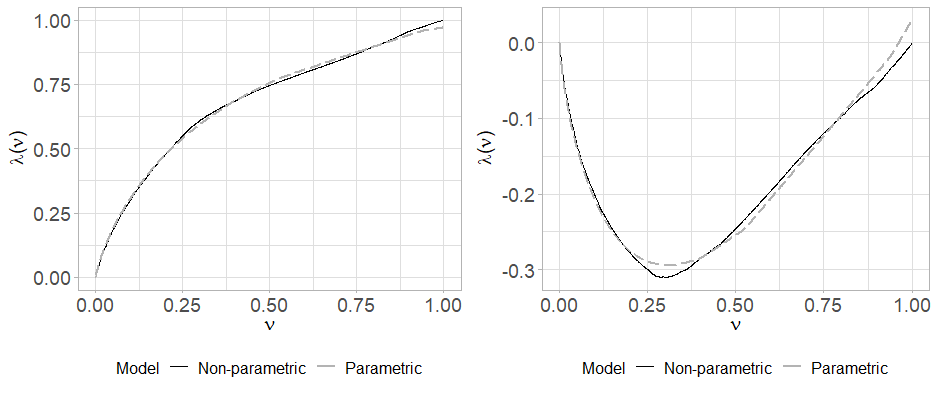}
\caption{$K(\nu)$ with Beran's estimator (black lines) and GAMLSS (blue dashed lines) for all data, with censored delays set as the settlement delays.}
\label{fig:coop_nocov}
\end{figure}

Figure \ref{fig:coop_nocov} shows the estimated Kendall distributions and corresponding $\lambda(.)$ functions with both approaches. Both give almost identical results in the shapes of dependence. The strenght of dependence, captured via the estimated Kendall's tau is also almost identical with both models. For the non-parametric model, with obtain $\hat{\tau} = 0.2534$ and for the parametric model, $\hat{\tau}=0.2569$.

\subsection{Dependence modeling with a covariate}
We now focus on the parametric model only and use $Z$ as the conditioning variable. We estimate four univariate Kendall distributions: one for the dataset as a whole, $\hat{K}(\nu)$, one for the subset of the data in which policyholders were born after 1990, $\hat{K}(\nu | Z = 1)$, one in which they were born between 1970 and 1989, $\hat{K}(\nu | Z = 2)$, and one in which they were born before 1969, $\hat{K}(\nu | Z = 3)$. We use again Weibull censored regressions for the marginal distributions of the activation delays. 

\begin{figure}[h!]
\centering
\includegraphics[width=1\textwidth]{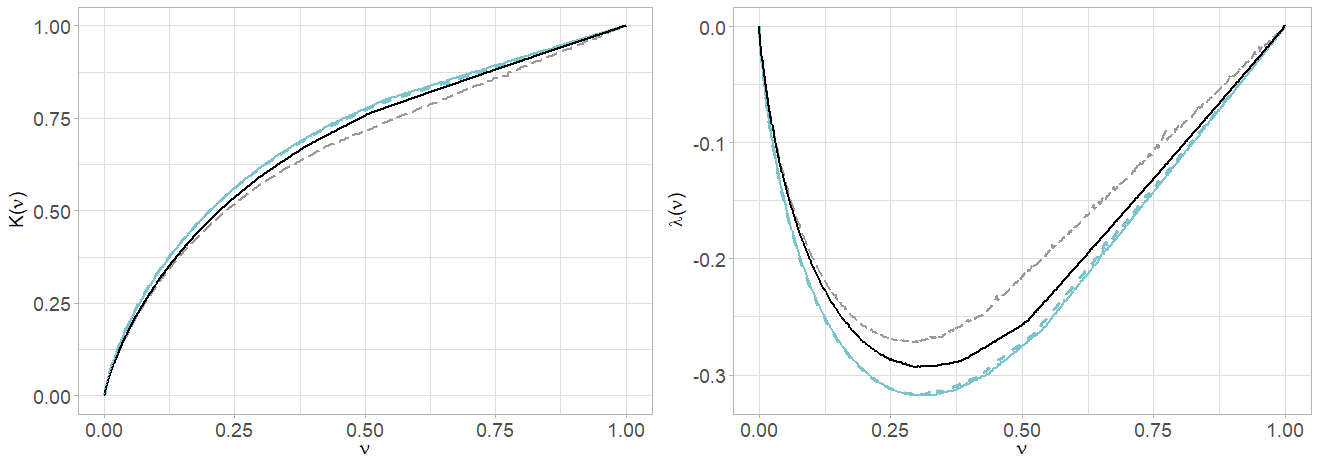}
\caption{$K(\nu)$ and $\lambda(\nu)$ when conditioning on $Z$.}
\label{fig:coop_covlevels}
\end{figure}

Figure \ref{fig:coop_covlevels} shows the plot of these four Kendall distribution curves on the left, and the corresponding curves for the $\lambda(.)$ function on the right. The black continuous curves depict, respectively, $\hat{K}(\nu)$ and $\hat{\lambda}(\nu)$, the dark grey dashed curves stand for $\hat{K}(\nu | Z = 2)$ and $\hat{\lambda}(\nu | Z = 2)$, the dark grey dotted curves represent $\hat{K}(\nu | Z = 1)$ and $\hat{\lambda}(\nu | Z = 1)$, and the light grey curves show $\hat{K}(\nu | Z = 3)$ and $\hat{\lambda}(\nu | Z = 3)$. The simplifying assumption clearly does not hold here, as different values of the age of policyholders result in different copulas. 

Table \ref{tab:coef} displays the estimated values of Kendall's tau for the data as a whole and for the different subsets. While conditioning on the age group of the policyholders does not appear to have a strong impact for the population as a whole, the results are quite different when we consider the different values of $Z$ separately. While the dependence seems to decrease for the younger and older policyholders in the portfolio, it increases from $25.81\%$ when considering all individuals to $33.8\%$ for the individuals born between 1970 and 1989. 

\begin{table}[h]
\begin{center}
\caption{Kendall's tau for different values of the covariate}
\label{tab:coef}
\begin{tabular}{@{}c c@{}}
\toprule
Model & $\hat{\tau}$ \\
\midrule
All &  0.2581 \\
$Z = 1$ & 0.1919 \\
$Z = 2$ & 0.3379 \\
$Z = 3$ & 0.1813 \\
\bottomrule
\end{tabular}
\end{center}
\end{table}

\begin{figure}[h!]
\centering
\includegraphics[width=1\textwidth]{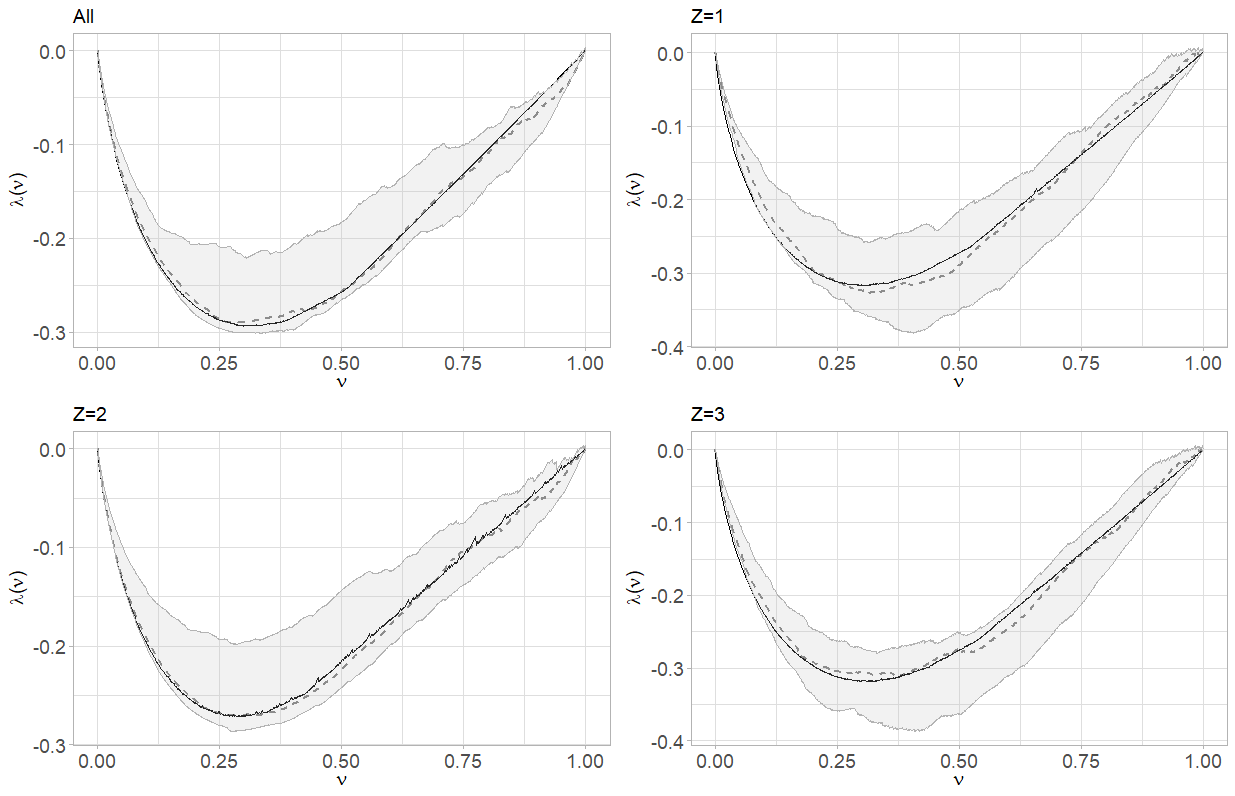}
\caption{$K(\nu)$ with Beran's estimator (black lines) and GAMLSS (blue dashed lines) for all data, with censored delays set as the settlement delays.}
\label{fig:coop_allsim}
\end{figure}

Next, we apply the same simulation procedure as the one described in Section \ref{sec:RDS} to investigate the best copula model in each scenario. For the complete population and for each of the three sub-groups of policyholders, we perform simulations from the estimators of the generator functions $\hat{\psi}(\nu)$, $\hat{\psi}(\nu | Z = 1)$, $\hat{\psi}(\nu | Z = 2)$ and $\hat{\psi}(\nu | Z = 3)$ using the \texttt{RLAPTRANS} algorithm. For each simulated sample, we select the most fitting model among the Clayton, Frank, Gumbel and Joe copulas by searching for the minimal $L^2$-norm. The graphical results of 1000 of these simulations are presented in Figure \ref{fig:coop_allsim}. In these plots, the black continuous curves are identical to those from Figure \ref{fig:coop_covlevels}, i.e. they represent the original $\hat{\lambda}(.)$ functions estimated from the data, for the whole population (top left) and when conditioning on the three different levels of the covariate $Z$. The grey dashed curves show the average of the simulations and the grey areas represent the $95\%$ confidence intervals. On average, using simulations from the generator functions in each case produces results that are very similar to the original data.  

The shapes of the curves in Figure \ref{fig:coop_allsim} seem to resemble each other. This indicates that, although the different age groups of policyholders clearly have an impact on the strength of dependence, the shape seems to remain stable. We see that it is not entirely true when looking at the results displayed in Table \ref{tab:L2}. For each conditioning scenario, this table shows the 1-$p$-values obtained from performing the simulation procedure described in Section \ref{sec:RDS} 1000 times. The copula model for which the displayed value is smallest is the one that was selected as best fitting by most simulations. We also present here the average estimate of Kendall's tau over all simulations for each conditioning scenario. 

For the population as a whole and for each of the three age groups, the Joe copula appears to be the best fitting model. For all policyholders, 35 out of 1000 simulations reject and Joe copula as best fitting model and opt instead for the Gumbel copula. For the policyholders for which $Z=2$, i.e. born between 1970 and 1989, only three of the thousand simulations reject the Joe model and favor the Gumbel copula. For the whole population and for this sub-group, none of the simulations select the Frank or Clayton copulas. The Joe model comes out as the largely favoured choice in these two cases. 

It is however not as evident for the younger and older policyholders. Although most simulations in both these cases still favor a Joe copula, there seems to be more variability in the choice. For the younger policyholders, that is those for whom $Z=1$, 384 simulations out of a thousand reject the Joe copula, i.e. ten times more than for the population as a whole and a hundred times more than for the policyholders born between 1970 and 1989. Among these 384 simulations, 344 opt instead for a Gumbel copula, 27 for a Frank copula and the remaining 13 for a Clayton model. For the older policyholders, 266 simulations reject the Joe model. They opt for the Gumbel (233), Clayton (20) and Frank copulas (13). Although the Joe copula still appears best suited for both sub-groups, almost $40\%$ and $27\%$ of simulations reject it for, respectively, the younger and older policyholders. This clearly shows that different levels of the covariate have an impact on the dependence structure of the data.  

\begin{table}[h!]
\begin{center}
\caption{Results from the copula selection procedure using the $L^2$-norm.}
\label{tab:L2}
\begin{tabular}{@{}c c c c c c@{}}
\toprule
$Z$ & Clayton & Frank & Gumbel & Joe & $\Bar{\tau}$ \\
\midrule
All & 1.000 & 1.000 & 0.965 & 0.035 & 0.2512\\
$Z=1$ & 0.987 & 0.973 & 0.656 & 0.384 & 0.1925\\
$Z=2$ & 1.000 & 1.000 & 0.997 & 0.003 & 0.3451\\
$Z=3$ & 0.980 & 0.987 & 0.767 & 0.266 & 0.1882\\
\bottomrule
\end{tabular}
\end{center}
\end{table}

Next, we illustrate the results of using the four different approaches described below to predict the activation delays for both coverages.
\begin{itemize}
    \item Approach 1: we do not condition on the different levels of $Z$ and use the results displayed in Table \ref{tab:L2}, that is a Joe copula with Kendall's tau set to $0.2512$, to perform predictions for the activation delays of the population as a whole.
    \item Approach 2: we condition on the three levels of the covariate and use again the results from Table \ref{tab:L2} to predict the activation delays for the different sub-groups of policyholders. We thereby use Joe copulas with Kendall's tau equal to $0.1925$, $0.3451$ and $0.1882$ for, respectively, the younger, middle-aged and older policyholders.
    \item Approach 3: we do not condition on the different levels of $Z$ and obtain predictions by simulating directly from the generator $\hat{\psi}(\nu)$ using the \texttt{RLAPTRANS} algorithm.  
    \item Approach 4: we condition on the three levels of $Z$ and obtain predictions by simulating directly from the three generators $\hat{\psi}(\nu | Z=1)$, $\hat{\psi}(\nu | Z=2)$ and $\hat{\psi}(\nu | Z=3)$, using again the \texttt{RLAPTRANS} algorithm.
\end{itemize}

\begin{figure}[h!]
\centering
\includegraphics[width=1\textwidth]{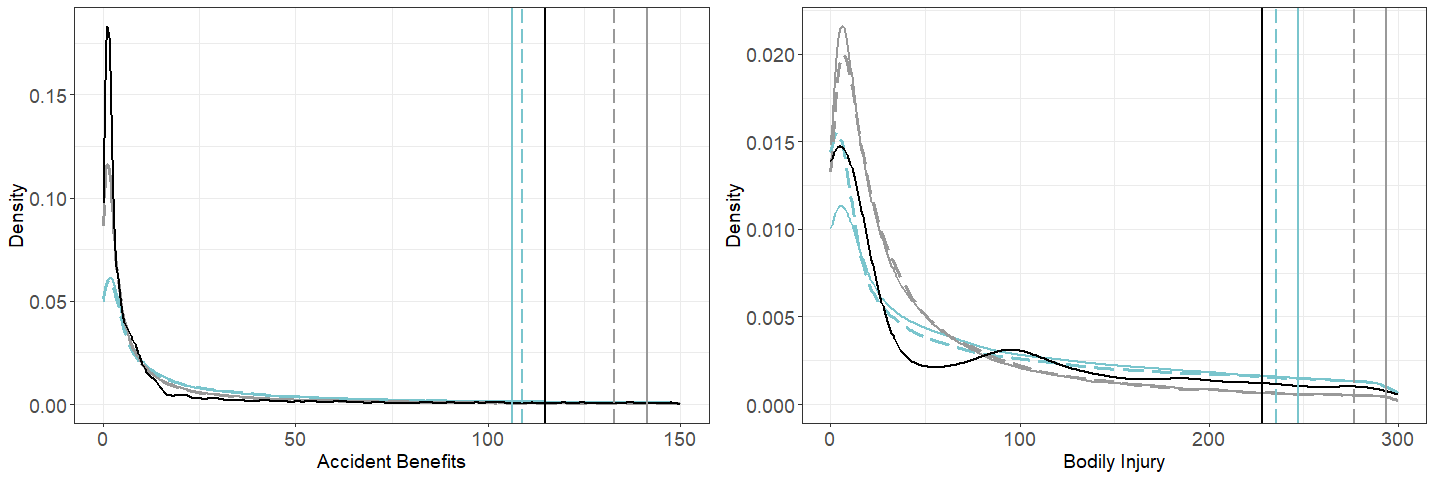}
\caption{Observed and simulated densities of the activation delays for both coverages, using the four different approaches.}
\label{fig:delayssim}
\end{figure}

While approaches 1 and 2 use known copulas, approaches 3 and 4 do not impose a specific model. Figure \ref{fig:delayssim} shows the simulated densities for the activation delays of the Accident Benefits (left) and Bodily Injury (right) coverages. On each plot, the black curves represent the observed densities. The grey continuous (resp. dashed) curves show the densities resulting from approach 1 (resp. approach 2) and the blue continuous (resp. dashed) curves display the densities resulting from approach 3 (resp. approach 4). The vertical lines represent the observed average activation delays (in black) and average simulated delays for the corresponding approaches. These values are also displayed in Table \ref{tab:simdelays}.

We first observe on the plots of Figure \ref{fig:delayssim} that the simulated densities of approaches 1 and 2 are quite similar and that the same goes for approaches 3 and 4. This is particularly true for the Accident Benefits coverage where the two grey (resp. blue) curves almost entirely overlap. For the Bodily Injury coverage, the blue curves, corresponding to the densities simulated directly from the generator functions, appear to be closer to the true density of the activation delays. 

We now look at the average values represented both by the vertical lines on the plots of Figure \ref{fig:delayssim} and by the values in Table \ref{tab:simdelays}. For the approaches using Joe copulas, namely approaches 1 and 2, conditioning on the age group of the policyholders appears to provide predictions that are closer to the observed activation delays. For the Accident Benefits coverage, approach 1 provides the highest prediction with an average delay of $141.47$ days, while when conditioning on the covariate $Z$ in approach 2, the average predicted delay decreases to $132.65$ days, closer to the observed $114.81$ days. We observe similar patterns for the Bodily Injury coverages. Approaches 3 and 4 show comparable results, although the difference between them is slightly less marked. For the Bodily Injury coverage, using the generator function for all policyholders provides an average estimated delay of $247.05$ days while using the generator functions for the three different age groups results in an average delay of $235.59$ days. For the Accident Benefits coverages, approach 3 results in an average estimated delay of $106.23$ days, while it is $108.86$ days for approach 4. 

 For both the Accident Benefits and Bodily Injury coverages, the observed activation delays are, respectively, $114.81$ and $228.39$ days. We can thereby conclude this section with the two following interesting observations. First, with both sets of approaches, namely approaches 1 and 2 based on the Joe copulas and approaches 3 and 4 based on the generator functions, the predictions closest to the true activation delays are obtained when conditioning on the age group of the policyholders. Second, using the generator functions rather than the Joe copulas and thereby not imposing a known model seems to lead to closer average predictions.

\begin{table}[h!]
\begin{center}
\caption{Average activation delays (observed and simulated).}
\label{tab:simdelays}
\begin{tabular}{@{}l c c c c c@{}}
\toprule
Coverage & Observed & Approach 1 & Approach 2 & Approach 3 & Approach 4  \\
 \midrule
 Accident Benefits  & 114.81 & 141.47 & 132.65 & 106.23 & 108.86\\
 Bodily Injury  & 228.39 & 293.66 & 276.75  & 247.05 & 235.59\\
 \bottomrule
\end{tabular}
\end{center}
\end{table}

\section{Conclusion}
\label{sec:conclusion}
In this paper, we present a new parametric estimator for the generator function of Archimedean copulas that is suitable for censored data. By incorporating covariates in the estimator, we allow both the strength and shape of dependence to vary with different values of the endogenous variables, thereby bypassing the simplifying assumption often used in conditional copula models. Our model allows to capture the effects of some covariates of interest simply by conditioning the joint distribution on them. We demonstrate its performance in two different applications. 

First, we study the diabetic retinopathy dataset and model the dependence between the times to blindness in both eyes for diabetic patients, as well as the effects of the age at onset of diabetes on this dependence. We do this for the patients population as a whole then investigate the model for different values of the covariate. In line with the results presented in \cite{geerdens2018}, we find that different ages at onset of diabetes lead to different estimates of Kendall's tau, the level of correlation being lower (resp. higher) for patients who were diagnosed with diabetes at a younger (resp. older) age. Even more interestingly, we show that in addition to impacting the strength of dependence, different values of the conditioning variable lead to different copula models. Although, because of high levels of censoring and a rather small sample of observations, we may not be able to define the copula families best suited to different levels of the covariate with high degrees of confidence, our model still allows us to use specific copulas for these different levels, thanks to the fact that we directly estimate the generator function rather than the copula itself.

Next, we provide an application to micro-level reserving. We use a Canadian automobile insurance dataset in which each policy in force provides multiple coverages to the insureds. We model the dependence between two of these coverages by means of their activation delays while conditioning on the age of the policyholders. We find again that different levels of the covariate provide different estimates of Kendall's tau. We also observe an impact of the covariate on the structure of dependence. Although Joe copulas are selected as best fitted to the data in both the unconditional and conditional models, this choice is not evident for some levels of the conditioning variable. We compare four different prediction approaches for the activation delays of the insurance coverages, using Joe copulas or the estimated generator functions, with and without conditioning on the age group of the policyholders. We show that using direct simulations from the generator function provides predictions that are closer to the observed activation delays. In addition, conditioning on the covariate also seems to result in better predictions, whether using Joe copulas or the generators. 

These applications demonstrate the benefits that can be reaped from considering the impact of endogenous variables in dependence models. By doing so, we seem to be able to better capture the specific dependence structure present in a dataset. In a claims reserving setting as the one described in Section \ref{sec:insurance} where a lot of information is available to insurers, we show that making use this information can lead to predictions that more closely fit the data.

In this paper, we have selected the levels of the conditioning covariates for which to analyse the dependence structure in a trivial way. For both the diabetic retinopathy study and the loss reserving application, we chose splitting points that provided subsets of similar size. It would be interesting, in future research, to explore a more robust method to select the levels of the conditioning variables at which the dependence model experiences a significant change. 

\clearpage







\newpage
\bibliographystyle{unsrt}  
\bibliography{main}  

\newpage
\appendix

\section{Simulating bivariate samples from a univariate distribution}
\label{app:1}

In this appendix, we provide more details on the Marshall, Olkin (\cite{marshall1988}) and \texttt{RLAPTRANS} (\cite{ridout2009}) algorithms that can be used to simulate a $d$-variate vector of dependent random variables from the generator function of Archimedean copulas. This strategy relies on Bernstein's theorem.
\begin{theorem}[Bernstein]
    A function $\psi(.)$ is strictly monotonic and $\psi(0)=1$ if and only if it can be written as 
    \begin{align}
        \psi(\nu) = \mathcal{L}_{\Theta}(\nu),
    \end{align}
    where $\mathcal{L}(.)$ is the Laplace-Stieltjes transform of the strictly positive random variable $\Theta$.
\end{theorem}

Knowing that Archimedean generators can be expressed as the Laplace-Stieltjes transform of a random variable $\Theta$, we can generate samples from this random variable and use an algorihtm such as the Marshall, Olkin algorithm described below to obtain a $d$-variate vector from an Archimedean copula characterized by its generator. 

\begin{algorithm}
\caption{Marshall, Olkin}\label{alg:MO}
\begin{algorithmic}[1]
        \item Generate a random observation $\theta$ from the distribution with Laplace transform $\psi$.
        \item For $i=1,...,d$, generate i.i.d. $X_i \sim \text{U}(0,1)$ .
        \item Return $(U_1,...,U_d)$ where $U_i = \psi(-\log(X_i)/\theta)$, for $i=1,...,d$.
\end{algorithmic}
\end{algorithm}

Algorithm \ref{alg:RLAP}, the so-called \texttt{RLAPTRANS} algorithm, provides a simple strategy based on a standard modification of the Newton-Rapjsom method for the first step of Algorihtm \ref{alg:MO}. Knowing the generator function $\psi(.)$, we can sample from its inverse Laplace-Stieltjes transform and generate observations from the random variable $\Theta$.

\begin{algorithm}
\caption{RLAPTRANS}\label{alg:RLAP}
\begin{algorithmic}[1]
    \item Generate $n$ independent $U(0,1)$ observations and sort them: $u_{(1)} < ... < u_{(n)}$.
    \item Find a value $\theta_{max}$ to serve as upper bound, i.e. $\psi^{-1}(\theta_{max}) \geq u_{(n)}$.
    \item Set the lower bound $\theta_L = \theta_{(i-1)}$ and upper bound $\theta_U = \theta_{max}$. Repeat the modified Newton-Raphson procedure for $i=1,...,n$ to obtain the ordered sample $\theta_{(1)},...,\theta_{(n)}$.
    \item Permute the ordered sample randomly to obtain the unordered sample $\theta_1,...,\theta_n$.
\end{algorithmic}
\end{algorithm}

\end{document}